# A Scalable Arrangement Method for Aperiodic Array Antennas to Reduce Peak Sidelobe Level

**Jiao Zhang[1], Hongtao Zhang[1], Xuelei Chen[2], Fengquan Wu[2], Yufeng Liu[1], Wenmei Zhang[1]**

[1]School of Physics and Electronic Engineering, Shanxi University, Taiyuan 030006, China
[2]National Astronomical Observatory, Chinese Academy of Science, Beijing 100101, China.

Corresponding author: Jiao Zhang (e-mail: zhangjiao@sxu.edu.cn).

This work was supported in part by the Chinese Academy of Sciences (CAS) instrument grant ZDKYYQ20200008; in part by the CAS Strategic Priority Research Program XDA15020200; and in part by the National Natural Science Foundation of China under Grants 11974228.

**ABSTRACT** Peak sidelobe level reduction (PSLR) is crucial in the application of large-scale array antenna, which directly determines the radiation performance of array antenna. We study the PSLR of subarray level aperiodic arrays and propose three array structures: dislocated subarrays with uniform elements (DSUE), uniform subarrays with random elements (USRE), dislocated subarrays with random elements (DSRE). To optimize the dislocation position of subarrays and random position of elements, the improved Bat algorithm (IBA) is applied. To draw the comparison of PSLR effect among these three array structures, we take three size of array antennas from small to large as examples to simulate and calculate the redundancy and peak sidelobe level (PSLL) of them. The results show that DSRE is the optimal array structure by analyzing the dislocation distance of subarray, scanning angle and applicable frequency. The proposed design method is a universal and scalable method, which is of great application value to the design of large-scale aperiodic array antenna.

## I. INTRODUCTION

In the design of array antenna, the reduction of PSLL is very important because it will directly affect the power characteristics and anti-interference performance of the system [1]. The PSLL, usually referring to the normalized maximum sidelobe level value, is the secondary lobe level in the radiation pattern. When the PSLL of array antenna is too high, it will affect the main lobe and cause interference and loss to the energy of the whole array antenna. At the same time, its echo will also interfere with the radar system. Therefore, in order to improve the overall performance of array antenna, it is necessary to reduce PSLL.

There are several ways to reduce PSLL in aperiodic array antennas. Elements optimization is the most basic method, which is realized by optimizing each array element in a completely random way or following some certain pattern. Smolders [2] has shown a concept of random sequential rotation to provide well-controlled sidelobes in a regular periodic array grid, which has similar ability as randomly-spaced array antennas. Lucas [3] has reported a Fermat's spiral based array antenna to suppress secondary radiation lobes, which has similar sidelobe level and bandwidth as those obtained by computationally expensive non-linear optimization methods. The aperiodic weighted array antennas proposed in literature [4] have excellent bandwidth characteristics which can effectively minimize the PSLL. Haupt [5] has used genetic algorithms by encoding parameters containing location information as binary strings, and the array distribution with the lowest maximum relative sidelobe level is obtained after finite iteration optimization. Due to the outstanding characteristics of aperiodic array antenna, such as wide bandwidth, high resolution and wide scanning angle, it has been widely used in satellite communication, remote sensing [6]-[8] and medical imaging [9], [10].

Aperiodic array antennas with optimized elements can greatly reduce PSLL, but for large and medium-sized array, the array structure with high degree of freedom makes it difficult to realize wave control and power feeding. An improved idea is to design subarray level aperiodic array antennas [11]-[14], which is a compromise between random array and periodic array. For large scale antennas, a group of antennas can be regarded as an element, namely a subarray, and these subarrays can be placed randomly to greatly reduce the difficulty of analysis. Due to the identity of subarray structures, subarray level aperiodic array [11] can reduce the complexity of design and manufacture, and its PSLR ability is comparable to that of random array.

Both literatures [11], [12] designed the aperiodic arrays with the combination of aperiodic subarray (dislocation) and uniform array elements. The subarray was integrated with multi-channel transceiver module with periodic active channel arrangement, which can greatly simplify the manufacturing technology and effectively reduce the highest sidelobe level. Literatures [13], [14] designed the aperiodic arrays, which is combined of aperiodic subarray (rotating) and non-uniform array elements. Kiersten [13] have studied an aperiodic array composed of aperiodic subarrays (rotating) and non-uniform elements, in which the bandwidth capability and the sidelobe level (SLL) of various arrays were characterized by probability, and their capability of low-cost applications was analyzed. The research shows that the array with rotating random subarrays had good antenna characteristics and the feasibility of low-cost manufacturing. Junming [14] have done some further studies on the array antenna with rotating random subarrays, such as, the number of array elements, the density of array elements and the PSLL, and have done some comparisons with those of pure random array and periodic array antenna, such as, the bandwidth, directivity and design calculation complexity, which further highlights the advantages of subarray-level aperiodic array in design and manufacture.

In this paper, we consider a scalable arrangement method for large and medium-sized subarray level aperiodic array antennas, and propose three aperiodic array structures: DSUE, USRE, and DSRE. The array structure is optimized by IBA to reduce PSLL. The characteristics of different arrays are analyzed by redundancy theory. The influences of the number of elements and subarrays, the dislocation distance of subarrays, scanning characteristics and applicable frequency width on PSLL are analyzed. The simulation results show that DSRE not only has the best PSLR effect, but also has lower computational complexity.

The structure of this paper is arranged as follows. Section II describes the structure of subarray level aperiodic array antennas. Section III introduces array synthesis, application details of the IBA and redundancy theory. Section IV analyzes the PSLR effect of the array in terms of the number of elements and subarrays, subarray dislocation distance, scanning characteristics and applicable frequency range. Section V contains the conclusion and discussion.

## II. SUBARRAY LEVEL APERIODIC ARRAY DESIGN
We propose three kinds of aperiodic rectangular grid array antennas, each of which is composed of the same subarrays. Fig. 1 is the analysis model diagram of three array structures when the number of elements in the subarray is $M \times M$ ($4 \times 4$ in the figure) and the number of subarrays is $N \times N$ ($7 \times 7$ in the figure). The whole array is composed of the same kind of subarrays. Three kinds of array antennas are formed,

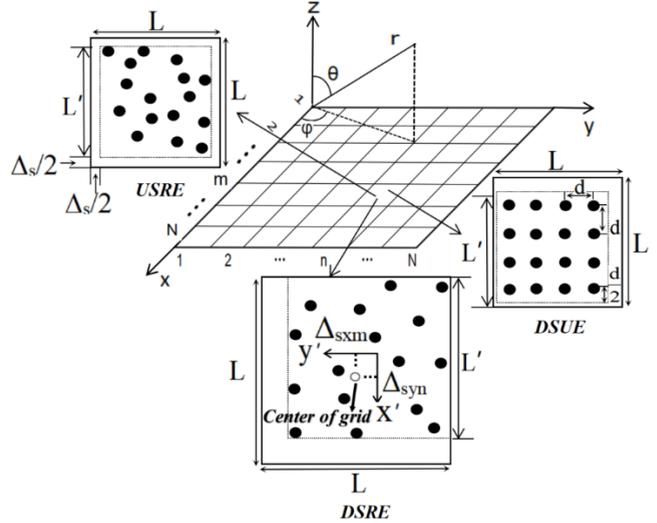

**FIGURE 1.** Analysis model diagram of three aperiodic array structures with relevant Cartesian (x,y,z) and spherical (r,θ,φ) coordinate systems. mn: the subarrays in the m-th row and n-th column; d: element spacing; L': The size of subarray; L: the size of grid; $\Delta_s$: the size of grid; $\Delta_s$: the longitudinal or transverse dislocation distance of a subarray; $\Delta_{sxm}$ and $\Delta_{syn}$: represent the dislocation distance of subarrays in x and y directions respectively.

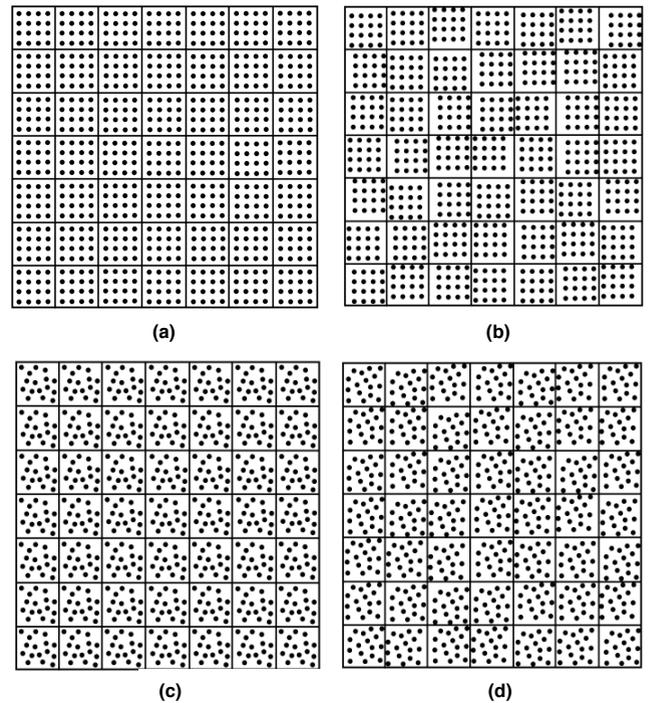

**FIGURE 2.** Structure diagrams of full arrays with 28×28 elements. The average array element spacing is slightly greater than one wavelength. (a) subarrays and array elements are uniform (b)DSUE (c) USRE (d)DSRE

namely, DSUE, USRE and DSRE. The size of the subarray is $L'$. The subarray is placed in a square grid with side length $L$ because of the need to vacate the dislocation distance of subarray. The whole array consists of squares grids of the same size which are uniformly and closely arranged, and the distance between the side of subarray and the same side of square grid is $\Delta_s/2$. $\Delta_s$ is the longitudinal or transverse dislocation distance of a subarray. There is no strict standard for its value, which needs to be determined according to the designed array. For the array with the average element spacing greater than one wavelength in this paper, $\Delta_s$ generally takes slightly less than one wavelength as the optimal value. The influence of dislocation position is briefly analyzed in the IV section.

Taking 28×28 array elements as an example, Fig. 2 (a) is the periodic array structure diagram with uniformly arranged array elements, and Fig. 2 (b), (c) and (d) are the array structure diagrams of DSUE, USRE and DSRE respectively. Among them, the structure of DSUE array has been discussed in literature [11]. The existing literature data show that USRE and DSRE arrays are rarely studied.

When the number of elements in the array is fixed, the number of subarrays is inversely proportional to the number of elements in the subarray. For the array with $28 \times 28$ elements, it can be divided into $7 \times 7$, $4 \times 4$ and $2 \times 2$ subarrays, and the corresponding number of elements in the subarray is $4 \times 4$, $7 \times 7$ and $14 \times 14$ respectively. The primary purpose of this paper is to compare the PSLR characteristics of these three arrays. It is very important to ensure the same resolution (related to array aperture) of arrays with different structures. At the same time, when changing the size of subarrays, the difference between array element spacing and subarray dislocation distance should not be too large. Therefore, we make them change proportionally, as shown in the following formula, and the array apertures are the same

$$\begin{cases} k = \frac{\Delta_{s7\times7}}{d_{4\times4}} = \frac{\Delta_{s4\times4}}{d_{7\times7}} = \frac{\Delta_{s2\times2}}{d_{14\times14}} \\ L_{a,7\times7} = L_{a,4\times4} = L_{a,2\times2} \end{cases} \quad (1)$$

where $k$ is the proportional constant, $\Delta_s$ is the subarray dislocation distance, $L_a$ is the array aperture, the corresponding subscript is the number of subarrays, $d$ is the element spacing, and its subscript is the number of elements in the subarray. For the random distribution of subarrays, the subarray length $L'$ of different arrays is also calculated by the uniform spacing of array elements.

## III. ANALYSIS METHOD

### A. ARRAY SYNTHESIS USING SUBARRAYS
The total radiation pattern function of the planar array can be expressed as [15]

$$F(\theta, \varphi) = f_e(\theta, \varphi) \, \text{AF}_{\text{array}}(\theta, \varphi) \quad (2)$$

where $f_e(\theta, \varphi)$ is the elemental pattern and $\text{AF}_{\text{array}}(\theta, \varphi)$ is the array factor of the planar array. Because the element radiation pattern changes slowly with the angle, and the PSLL reduction level of the array radiation pattern is close to that of the array factor, only the array factor is analyzed in this paper. According to the analysis model of array structure in Fig. 1, the array factor can be expanded as follows

$$\text{AF}_{\text{array}}(\theta, \varphi) = \sum_{m,n} \sum_{m'n'} I_{mn,m'n'} \, e^{jk[(s_{xm} + e_{xm'}) u_x + (s_{yn} + e_{yn'}) u_y]} \quad (3)$$

where,

$$\begin{aligned} s_{xm} &= x_m + \Delta_{sxm} \\ s_{yn} &= y_n + \Delta_{syn} \\ e_{xm'} &= m'x_o + \Delta_{x,m'n'} \\ e_{yn'} &= n'y_o + \Delta_{y,m'n'} \\ u_x &= \sin\theta\cos\varphi - \sin\theta_o\cos\varphi_o \\ u_y &= \sin\theta\sin\varphi - \sin\theta_o\sin\varphi_o \\ u &= \sin\theta\cos\varphi, \quad v = \sin\theta\cos\varphi \end{aligned} \quad (4)$$

where $mn$ and $m'n'$ represent the $mn$-th array element in the $m'n'$-th grid. To simplify the analysis, $I_{mn,m'n'}$ is the unit excitation amplitude, and its value is 1. $\Delta_{sxm}$ and $\Delta_{syn}$ represent the dislocation distance of the subarray in the $x$ and $y$ directions respectively. $(x_m, y_n)$ is the grid center coordinate. $(s_{xm}, s_{yn})$ represents the position of the subarray in the x and y directions, and $(x_o, y_o)$ is the position of the uniform subarray element. $(\Delta_{x,m'n'}, \Delta_{y,m'n'})$ is the offset distance from $(x_o, y_o)$. $(e_{xm'}, e_{yn'})$ represents the position of the array element in the $x'$ and $y'$ directions in the subarray. $(\theta_o, \varphi_o)$ is the maximum direction of the radiation pattern.

If all subarrays in the entire array have the same structure, equation (3) can be derived as the product of two factors, as shown in equation (5)

$$\begin{aligned} \text{AF}_{\text{array}}(\theta, \varphi) &= \sum_{m,n} e^{jk[s_{xm} u_x + s_{yn} u_y]} \times \sum_{m'n'} e^{jk[e_{xm'} u_x + e_{yn'} u_y]} \\ &= \text{AF}_{\text{ps}}(\theta, \varphi) \times \text{AF}_s(\theta, \varphi) \end{aligned} \quad (5)$$

where $\text{AF}_{\text{ps}}(\theta, \varphi)$, $\text{AF}_s(\theta, \varphi)$ are the subarray position factor and subarray factor.

Then the PSLL can be calculated by

$$\text{PSLL(dB)} = 20\log_{10}\left(\frac{\text{AF}_{\text{sl,max}}(u,v)}{\text{AF}_{\text{main}}}\right) \quad (6)$$

where $\text{AF}_{\text{sl,max}}(u,v)$ is the maximum sidelobe value of the radiation pattern. $\text{AF}_{\text{main}}$ is the peak value of the main lobe.

### B. IMPROVED BAT ALGORITHM
With the improvement of computing capability, many heuristic intelligent optimization algorithms have been used to optimize complex, nonlinear, and irregular array factors, such as genetic algorithm [16], particle swarm optimization [17], firefly algorithm [18], differential evolution algorithm [19], and super multivariate optimization algorithm applied to large array optimization [20]. Bat algorithm [21] [22] is a heuristic search algorithm based on swarm intelligence, which is more suitable for solving multi-dimensional and multi-constraint optimization problems. Its good characteristics have been affirmed in the non-uniform linear array [23].

In order to analyze the array structure, we upgrade the IBA, which is proposed in our previous work [21]. This IBA is used to reduce the PSLL by optimizing the position of multiple constrained antenna array. More complicated constraints such as array spacing and array aperture will be considered in the optimization of the algorithm.

The steps and parameters of IBA are as follows:

*Step1*: Initialize the population and parameters. The initial population can be increased to improve the accuracy of the initial solution. On the premise that the initial position solution is reliable, the flight speed $v_i$ of the bat should be appropriately reduced at this position. It prevents flying too fast from leaving the optimal solution range. Table 1 shows the specific parameter settings.

TABLE I
PARAMETER SETTING OF IMPROVED BAT ALGORITHM

| Parameter | Value |
|---|---|
| Maximum number of iterations | 50 |
| Population size | 300 |
| Pulse rate factor $\gamma$ | 0.9 |
| Frequency $f$ | 0-1 |
| Initial Loudness $A$ | 1-2 |
| Attenuation factor $\alpha$ | 0.9 |
| Flight speed $v$ | 0-0.5 |
| Inertia weight $\omega$ | 0.5-0.9 |
| Doppler effect compensation rate $C$ | 0.1-0.9 |

When the initial population is generated, it is necessary to judge the array element spacing of the initial population and modify the position of the array elements that do not meet the constraint conditions. The standard for judging the array element spacing is the minimum array element spacing $d_{min} \geq \lambda/2$, and the settings of multiple constraint conditions are as follows

$$\begin{cases} s.t \quad d_{min} \leq d_{s,m'n'} - d_{s,m''n''} \leq d_{max} \\ -\Delta_s/2 \leq \Delta_{s,mn} \leq \Delta_s/2 \\ x_{a,mn} \leq L_a, \ y_{a,mn} \leq L_a, \ d_{a,mn} \leq 2L_a^{1/2} \end{cases} \quad (7)$$

where $\Delta_{s,mn}$ is the dislocation distance of subarray. $x_{a,mn}, y_{a,mn}$ indicate the constraint size of the total array in the $x$ and $y$ directions, $L_a = N(Md + \Delta_s)$ is the array aperture, and $d_{a,mn}$ is the spacing of elements in the array.

Under the condition of satisfying the array aperture and the array element spacing, the constraint spacing of elements in the subarray can be expressed as

$$\begin{cases} d_{s,m'n'} - d_{s,m''n''} = \left[ (x_{s,m'n'} - x_{s,m''n''})^2 + (y_{s,m'n'} - y_{s,m''n''})^2 \right]^{\frac{1}{2}} \\ d_{min} = \frac{\lambda}{2}, \quad d_{max} = 2L_a^{1/2} \end{cases} \quad (8)$$

where $x_{s,mn}, y_{s,mn}$ represent the position of elements in the subarray in the $x$ and $y$ directions, and $d_{s,mn}$ denotes the spacing of any two elements in the subarray.

When all array elements meet the constraint conditions, the initial population is generated and the initial solution is obtained.

*Step2*: Set the fitness function, substitute the initial population into the fitness function, and start iteration to generate a preliminary global optimal position solution. The fitness function is expressed as follows

$$f(A) = \min\{\text{fitness}(A)\} = \max \left\{ \frac{\text{AF}_{sl}(u,v)}{\text{AF}_{main}} \right\} \quad (9)$$

where $\text{AF}_{sl}(u, v)$ is the sidelobe area of radiation pattern, and the objective function will locate the PSLL in the whole airspace.

If a bat finds the best foraging position $g_*$, it will attract other bats to fly in search of food. Each bat is associated with a velocity $v_i^t$ and a position $x_i^t$ at iteration $t$. To update velocities and positions of all bats, Equations (10)-(13) are used.

$$f_i = f_{min} + (f_{max} - f_{min}) \times rand(0,1) \quad (10)$$

$$f_i' = \frac{c + v_i^{t-1}}{c + v_g} \times f_i \times \left(1 + C_i \times \frac{g_* - x_i^{t-1}}{|g_* - x_i^{t-1}| + \varepsilon}\right) \quad (11)$$

$$v_i^t = \omega \times v_i^{t-1} + \left(x_i^{t-1} - g_*\right) \times f_i' \quad (12)$$

$$x_i^t = x_i^{t-1} + v_i^t \quad (13)$$

where $f_{min}$ and $f_{max}$ are the minimum and the maximum frequency. Their values are 0 and 1, respectively, depending on the specific environment. $c = 340 \ m/s$ is the speed of sound in the air. $C_i$ denotes the doppler effect compensation rate. $\omega$ is the inertia weight.

Literature [21] has shown that the adaptive local search strategy with increased Doppler effect and Doppler effect compensation rate has a good effect of increasing population diversity. This paper also introduces it to improve the performance of the algorithm.

*Step3*: For the local search part, once a solution is selected among the current optimal solutions, a new solution is generated using a random walk

$$x_{new} = x_{old} + N[0,1] \times A^t \quad (14)$$

where $A^t$ is the average loudness of all bats. The new solution should also meet the constraint conditions.

*Step4*: The fitness function is used to evaluate the local new solution. When the new solution is better than the current optimal solution, the loudness and pulse emission rate are updated by the following equations

$$A_i^{t+1} = \alpha A_i^t \quad (15)$$

$$r_i^{t+1} = r_i^0 [1 - \exp(-\gamma t)] \quad (16)$$

where $\alpha$ and $\gamma$ are attenuation factor and pulse rate factor. The initial loudness $A_i$ can be [1,2], while the initial emission rate $r_i^0$ can be [0,1].

*Step5*: Update the current global optimal solution. If the loop reaches the maximum number of iterations, the optimal solution and the optimal fitness function value are output. If not, it returns to step2 to continue the iteration.

## C. REDUNDANCY THEORY

The minimum redundant array (MRA) is widely used in microwave radiometer [24], [25], radio astronomy [26], adaptive beamforming [27] and MIMO radar [28]. Any antenna pair in the array will form a baseline vector, and redundancy means that the baseline vectors are the same. The process of searching for MRA is to find the array structure with minimum redundancy.

Theoretically, a planar array with $N$ elements can have up to $N(N-1)$ non-redundant baselines. However, the actual number of non-redundant baselines is usually less than this value because the planar array elements are difficult to be completely random. Generally, the redundancy of an array is measured by $R$, which is defined as the ratio of the ideal number of non-redundant baselines $S_{id}=N(N-1)$ to the actual number of non-redundant baselines $S_{re}$.

The randomness of array element arrangement in aperiodic array is closely related to the PLSR effect. The redundancy can represent the uniformity of the array, and we can infer that the redundancy of the array can reflect its PSLR effect to some extent. It is not a mature theory to reduce PSLL by optimizing array redundancy, but the close mapping relationship between redundancy and aperiodic arrays with different uniformity can be seen through simulation in this paper.

## IV. SIMULATION RESULTS AND ANALYSIS

### A. THREE KINDS OF ARRAYS PSLR CHARACTERISTIC

In order to simplify the analysis, we take arrays with 12×12, 18×18 and 28×28 elements as examples. Three kinds of subarrays can be found for each case. For example, an array with 28×28 elements can be divided into 2×2, 4×4 and 7×7 subarrays. The following analysis methods and conclusions can generally be extended to other similar arrays with different numbers of array elements.

For all arrays in this paper, the positions of array elements are optimized by IBA at the frequency of 10 GHz. The average spacing of array elements is $d_{ave} \geq \lambda$. Fig. 3(a) is the comparison diagram of PSLR effect of three arrays which have 6×6 (array elements 12×12), 6×6 (array elements 18×18) and 7×7 (array elements 28×28) subarrays. The dislocation distance between these subarrays is $\Delta_s = 0.87\lambda$. When the number of subarrays is large, the PSLR effect of DSUE arrays is better than that of USRE arrays. This is because the dislocations of DSUE arrays have more non-uniformity than the randomness of USRE arrays when the number of elements in subarrays is relatively small. DSRE arrays can greatly break the periodicity of arrays, and their PSLR effect is superior to other arrays.

Fig. 3(b) and Fig. 3(c) illustrate the PSLR effect of three arrays when the numbers of subarrays are moderate and small, respectively. In order to ensure the same array aperture for the same number of array elements, the relationship between dislocation distance and element spacing should satisfy equation (1). In the case of 28×28 elements, the dislocation distances are $\Delta_{s,4\times4} = 0.93\lambda$ and $\Delta_{s,2\times2} = 0.99\lambda$, respectively. When the numbers of subarrays are less than the numbers of elements in the corresponding subarrays, the PSLR effect of USRE arrays is better than that of DSUE arrays, and that of DSRE arrays is still the best.

Fig. 4 shows the two-dimensional radiation patterns of the array structures corresponding to Fig. 2 sequentially. As shown in Fig. 4(a), the GL has the same amplitude as the main lobe. The peak sidelobe levels (PSLLs) and positions are also indicated in the radiation patterns. It can be clearly seen that due to the non-uniform arrangement of array elements, the peak sidelobe positions are distributed in the visible range of the entire airspace.

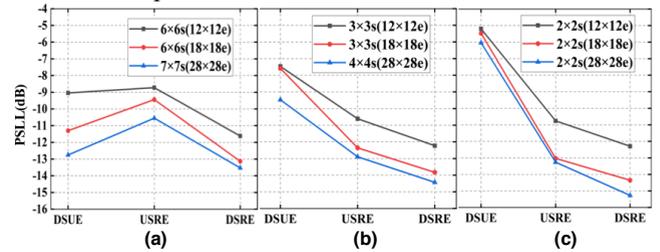

FIGURE 3. The PSLR effect of three arrays with different number of elements when the number of subarrays is (a) large; (b) moderate; (c) small. The s represents the number of subarrays. The e represents the number of elements in the array.

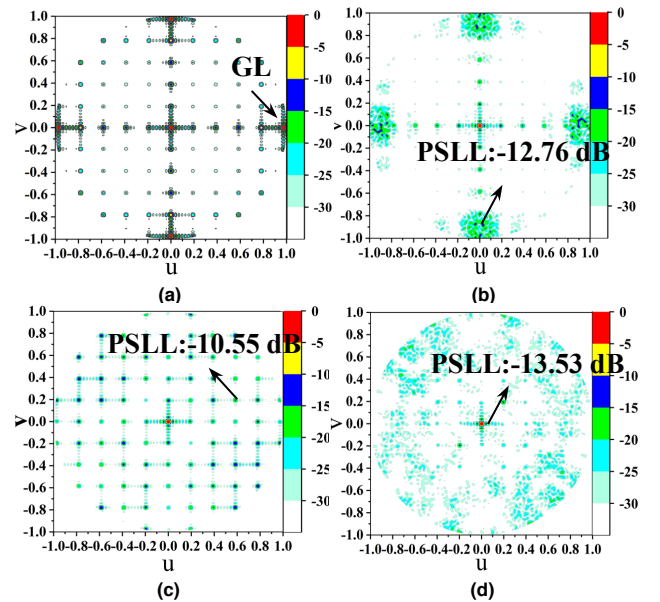

FIGURE 4. Two-dimensional radiation patterns corresponding to the array structure diagrams of Fig. 2. (a) subarrays and array elements are uniform (b)DSUE (c) USRE (d)DSRE

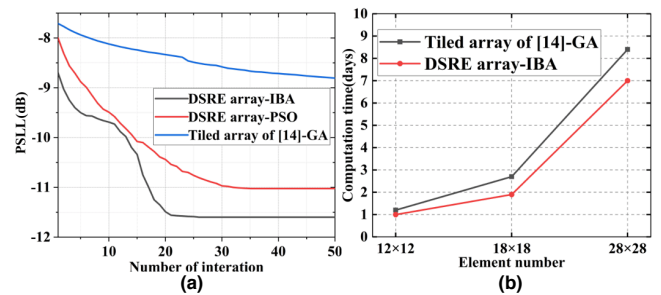

FIGURE 5. Optimization results for array antennas with 12×12 elements. (a) Convergence curves over iteration for IBA in this paper, traditional PSO and GA in literature [14]. (b) Comparison of calculation time over element number for IBA in this paper and GA in literature [14].

The corresponding relationship between PSLL and redundancy of arrays with different number of array elements is shown in Tables 2, 3 and 4. As can be seen from these tables, the three kinds of aperiodic arrays with different subarray divisions have different non-uniformity. The array structure with high degree of non-uniformity has less redundancy. Redundancy can effectively reflect the non-uniformity of arrays, which is helpful for researchers to grasp the randomness of arrays in the design process.

In literature [14], the array with rotating subarrays composed of 144 elements is optimized by traditional genetic algorithm (GA) at the frequency of 10 GHz. To highlight the proficiency of IBA, particle swarm optimization (PSO) is also selected to optimize DSRE$_{4×4}$ array with 12×12 elements. The initial population number is set to 200. When the average relative change in fitness function values over 10 generations is less than $1× 10^{-4}$, the optimization process is terminated. The optimization results of IBA are obtained when $v$=0.4, $\omega$=0.75. The PSO optimization results are calculated when the inertia constant $\omega = 0.7$, the acceleration constant $c_1$=$c_2$=1.5 and the maximum particle velocity $Vmax$ =0.12. As shown in Fig. 5(a), the convergence curves of these two algorithms are compared with those of GA applied to tiled array in literature [14]. It can be seen that IBA is superior to the other two algorithms in terms of convergence speed and accuracy. In this paper, IBA is calculated by Intel core i7 8700 processor. Compared with literature [14], the calculation time of IBA over element number is shorter, as shown in Fig. 5(b), which is closely related to the number of iterations and convergence characteristics of the algorithm.

As shown in Table 2, the PSLL of DSRE arrays with 12×12

elements worsen as the number of subarrays increases. The array length in literature [14] is 8$\lambda$, and the average array spacing is less than 1$\lambda$. The results show that the PSLL is about -11 dB. In this paper, in the case of the array length greater than 8$\lambda$ and the average array spacing slightly greater than 1$\lambda$, the worst PSLL is -11.62 dB, which is the result of PSLR with the largest number of subarrays (4×4). When the number of subarrays is 2×2, the PSLL can be suppressed to -12.29 dB. This is obviously superior to the subarray rotation structure proposed in literature [14].

As the number of array elements increases, the optimization time does not increase linearly, which depends on the accuracy of the initial population location of the algorithm. Once the population is located near the optimal position, the fitness function will converge rapidly, thus shortening the optimization time. According to the subarray division scheme, the optimization of array elements arrangement in each subarray only needs to be calculated once, and the dislocation optimization of all subarrays needs to be considered. Thus, the number of design variables is less than that of completely random arrays. Therefore, no matter the optimization time or the number of design variables, the DSRE array has low computational complexity.

In summary, the PSLR effect of DSUE and USRE arrays varies with the number of subarrays. No matter for which array structure, DSRE array is the best choice in the array design with higher requirements for PLSR, even for a small number of random elements in the subarray. Compared with other aperiodic arrays, such as subarray rotating random array, DSRE array also has good performance.

TABLE.2
CORRESPONDENCE BETWEEN PSLL AND REDUNDANCY OF ARRAYS WITH 12×12 ELEMENTS

| Number of array elements | 12×12 | | | | | | | | |
|---|---|---|---|---|---|---|---|---|---|
| Number of subarrays | 2×2 | | | 3×3 | | | 4×4 | | |
| Array structure | DSRE | USRE | DSUE | DSRE | USRE | DSUE | DSRE | USRE | DSUE |
| PSLL (dB) | -12.29 | -10.76 | -5.20 | -12.23 | -10.61 | -7.47 | -11.62 | -8.72 | -9.04 |
| Number of non-redundant baselines | 18616 | 15114 | 4758 | 18326 | 10008 | 5252 | 18044 | 5682 | 6998 |
| Redundancy ($R$) | 1.11 | 1.36 | 4.33 | 1.12 | 2.06 | 3.92 | 1.14 | 3.62 | 2.94 |

TABLE.3
CORRESPONDENCE BETWEEN PSLL AND REDUNDANCY OF ARRAYS WITH 18×18 ELEMENTS

| Number of array elements | 18×18 | | | | | | | | |
|---|---|---|---|---|---|---|---|---|---|
| Number of subarrays | 2×2 | | | 3×3 | | | 6×6 | | |
| Array structure | DSRE | USRE | DSUE | DSRE | USRE | DSUE | DSRE | USRE | DSUE |
| PSLL (dB) | -14.35 | -13.03 | -5.47 | -13.83 | -12.36 | -7.35 | -13.13 | -9.44 | -11.30 |
| Number of non-redundant baselines | 96938 | 79844 | 12854 | 96848 | 52518 | 13748 | 95222 | 24678 | 40624 |
| Redundancy ($R$) | 1.08 | 1.31 | 8.14 | 1.08 | 1.99 | 7.61 | 1.10 | 4.24 | 2.58 |

TABLE.4
CORRESPONDENCE BETWEEN PSLL AND REDUNDANCY OF ARRAYS WITH 28×28 ELEMENTS

| Number of array elements | 28×28 | | | | | | | | |
|---|---|---|---|---|---|---|---|---|---|
| Number of subarrays | 2×2 | | | 4×4 | | | 7×7 | | |
| Array structure | DSRE | USRE | DSUE | DSRE | USRE | DSUE | DSRE | USRE | DSUE |
| PSLL (dB) | -15.27 | -13.26 | -6.04 | -14.43 | -12.89 | -9.47 | -13.53 | -10.55 | -12.76 |
| Number of non-redundant baselines | 607620 | 460218 | 42574 | 584884 | 250328 | 83726 | 576616 | 118748 | 212452 |
| Redundancy ($R$) | 1.01 | 1.33 | 14.42 | 1.05 | 2.45 | 7.33 | 1.06 | 5.17 | 2.89 |

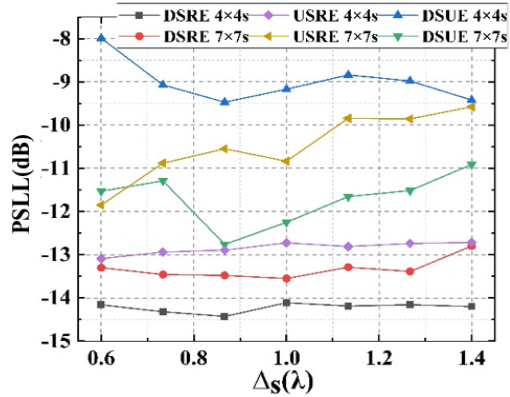

**FIGURE 6.** Influence of subarray dislocation distance on PSLLs.

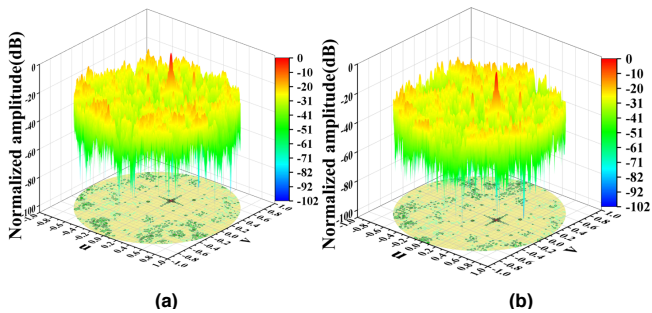

**(a)**        **(b)**

**FIGURE 7.** 3D scanning patterns and plane maps of DSRE array with 7×7 subarrays (a) u = 15° (b) v = 15° .

## B. INFLUENCE OF SUBARRAY DISLOCATION DISTAN-CE ON PSLR

It is very important to select the optimal dislocation distance of subarrays to improve the performance of array antennas in the subarray level aperiodic array. This section will mainly study the influence of dislocation distance on the PLSR effect. The array with 28×28 elements is still considered. From the perspective of engineering application, there are too many randomly arranged elements in the 2×2 subarrays, and the design of a single subarray is relatively complex and has low application value. So, it will not be analyzed here. This section discusses the PSLR effect of three kind of array structures under different dislocation distances with 7×7 and 4×4 subarrays.

In Fig. 6, the subarray dislocation distance is $0.6\lambda \leq \Delta_s \leq 1.4\lambda$. The array aperture is the same under the same subarray dislocation distance. The average array element spacing is no longer proportional to the subarray dislocation distance, but is determined by the array aperture.

According to the overall trend in Fig. 6, the PSLR effect is DSRE₄×₄>DSRE₇×₇>USRE₄×₄>DSUE₇×₇>USRE₇×₇>DSUE₄×₄. For the proposed structures, the PSLR effect is the best when the subarray dislocation distance is slightly less than a wavelength or the average array element spacing.

The PSLLs of DSRE₄×₄, DSRE₇×₇ and USRE₄×₄ arrays show a stable trend with the increase of $\Delta_s$. As shown in Table 4, the redundancy of these three arrays increases gradually. From the analysis of array structures, the non-uniformity of DSRE₄×₄ and DSRE₇×₇ arrays are affected by subarray dislocation and the non-uniform arrangement of array elements, with the latter being the dominant factor affecting the non-uniformity of array. Therefore, the more the array elements are arranged non-uniformly, the stronger the PSLR ability becomes. For USRE₄×₄ and DSUE₇×₇, the former has 7×7 non-uniformly arranged array elements, while the latter has 7×7 dislocated subarrays. The number of non-uniform influencing factors is the same, but the PSLR ability of the former is better than that of the latter. The reason is that the non-uniform arrangement of array elements plays a main role in influencing the non-uniformity of array. The same is true for USRE₇×₇ and DSUE₄×₄. When the number of single non-uniform influencing factors is different, such as USRE₇×₇ and DSUE₇×₇, the former has 4×4 non-uniformly arranged array elements, and the latter has 7×7 dislocated subarrays. In this case, the larger number of dislocated subarrays dominate the non-uniformity of the array, so the PSLR effect of the latter is better than that of the former.

The PSLR effect of all array structures in Fig. 6 basically weaken with the increase of $\Delta_s$, especially USRE₇×₇ and DSUE₇×₇. For arrays with a single non-uniform factor, it will inevitably lead to the increase of the subarrays spacing and the elements spacing with the increase of $\Delta_s$. The increase of either of them will worsen the PSLL of the array pattern. We can see that DSUE₄×₄ has a different change rule. The reason is that the number of dislocated subarrays is quite small, which is the only factor affecting the non-uniformity of the array. This results in its unspecific variation rule with $\Delta_s$.

## C. SCAN OF ARRAY

The scanning characteristic is an important index to evaluate the performance of array antenna. The phased array antenna with wide-angle scanning capability is widely used in radar, direction finding and remote sensing. This section will further verify the wide scanning characteristic of the array structures proposed in this paper.

For the array antenna with 28×28 elements, Fig. 7 is the full-space three-dimensional scanning patterns and plane maps of DSRE array with 7×7 subarrays when scanning angles are 15° in the u axis and 15° in the v axis. For a small scanning angle of 15°, the PSLL of the array radiation pattern is below -13.5 dB, and the SLL does not increase obviously. Fig. 8 shows the full-space three-dimensional scanning patterns and plane maps of DSUE array and USRE array with 7×7 subarrays when scanning angles are 15° in the u axis and 15° in the v axis. The u-axis and the v-axis can refer to the labels in Fig. 7. Table 5 lists the corresponding PSLL values in Fig. 7 and 8. Compared with the radiation pattern of DSRE array, it can be found that the scanning characteristics of DSUE array and USRE array are poor even at small scanning angles. With the increase of the scanning angle, the scanning characteristics are even worse, because the radiation energy is concentrated in some fixed areas, resulting in the increase of PSLL. However, due to the high non-uniformity of array elements, DSRE array can effectively disperse the radiation energy, and thus reduce the PSLL.

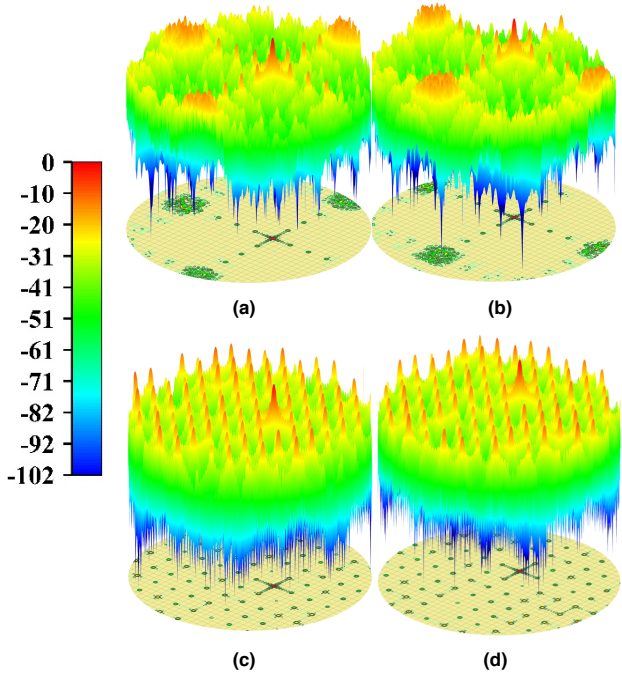

**FIGURE 8.** 3D scanning patterns and plane maps of DSUE array and USRE array with 7×7 subarrays (a) DSUE, *u* = 15° (b)DSUE, *v* = 15° (c) USRE, *u* = 15° (d)USRE, *v* = 15°.

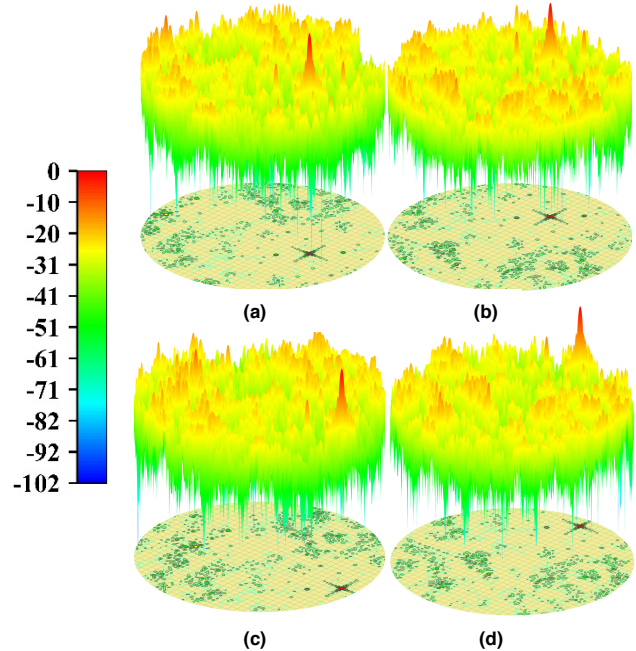

**FIGURE 9.** The 3D radiation patterns and plane maps of DSRE array with 7×7 subarrays scanned at 30° and 60° in the *u* and *v* axis. (a) *u* = 30° (b) *v* = 30° (c) *u* = 60° (d) *v* = 60°.

For a larger scanning range, DSRE array can also maintain good radiation characteristics. Fig. 9 shows the scanning patterns and plane maps at 30° and 60° in *u* axis and *v* axis respectively. And their PSLL values are shown in Table 6. It can be seen from the figure that at larger scanning angles of 30° and 60°, the PSLL and the average SLL do not increase significantly with the increase of scanning angle.

TABLE.5
PSLL OF THREE ARRAYS FOR DIFFERENT SCANNING ANGLES

| Array structure | DSRE | | DSUE | | USRE | |
|---|---|---|---|---|---|---|
| Scanning angles in *u* axis | 15° | 0° | 15° | 0° | 15° | 0° |
| Scanning angles in *v* axis | 0° | 15° | 0° | 15° | 0° | 15° |
| PSLL (dB) | -13.51 | -13.50 | -12.75 | -12.74 | -8.88 | -8.87 |

TABLE.6
PSLL OF DSRE ARRAY FOR DIFFERENT SCANNING ANGLES

| Array structure | DSRE | | | |
|---|---|---|---|---|
| Scanning angles in *u* axis | 30° | 0° | 60° | 0° |
| Scanning angles in *v* axis | 0° | 30° | 0° | 60° |
| PSLL (dB) | -13.45 | -13.43 | -13.05 | -13.04 |

TABLE.7
PSLL OF 7×7 SUBARRAYS FOR DIFFERENT FREQUENCIES

| Frequency (GHz) | 6.12 | 8.11 | 10.00 | 12.00 | 14.29 | 15.79 | 17.65 |
|---|---|---|---|---|---|---|---|
| PSLL(dB) 7×7s DSUE | -7.17 | -12.94 | -12.76 | -12.72 | -9.74 | -9.71 | -9.82 |
| PSLL(dB) 7×7s USRE | -5.24 | -10.49 | -10.55 | -8.88 | -7.66 | -5.25 | -5.14 |
| PSLL(dB) 7×7s DSRE | -7.28 | -13.51 | -13.53 | -13.49 | -10.12 | -10.14 | -10.13 |

TABLE.8
PSLL OF 4×4 SUBARRAYS FOR DIFFERENT FREQUENCIES

| Frequency (GHz) | 6.12 | 8.11 | 10.00 | 12.00 | 14.29 | 15.79 | 17.65 |
|---|---|---|---|---|---|---|---|
| PSLL(dB) 4×4s DSUE | -7.04 | -12.56 | -9.47 | -9.51 | -6.05 | -5.97 | -6.14 |
| PSLL(dB) 4×4s USRE | -7.18 | -12.90 | -12.89 | -9.74 | -9.77 | -9.81 | -9.87 |
| PSLL(dB) 4×4s DSRE | -7.21 | -14.32 | -14.43 | -14.34 | -14.42 | -14.54 | -14.39 |

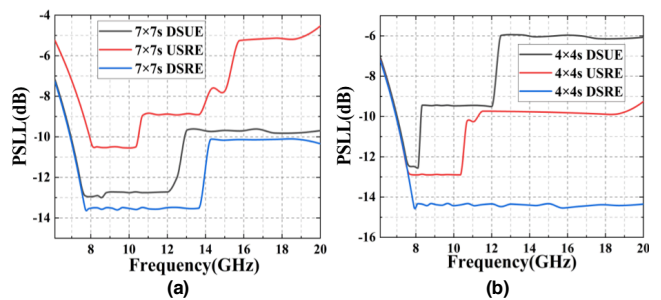

**FIGURE 10.** The PSLR effect of three structures with 28×28 elements in the frequency range (6GHz-20GHz). (a) 7×7 subarrays (b) 4×4 subarrays.

## D. FREQUENCY CHARACTERISTIC
Aperiodic array antennas have very stable modes over a wide frequency range, making them very suitable for operation in large bandwidths. This section will analyze the PSLR effect of three array structures at different frequencies and provide some design suggestions.

Fig. 10 shows the PSLR effect varying with frequency for 28×28 elements with 7×7 and 4×4 subarrays respectively. The array structures are optimized at the frequency of 10 GHz. Table 7 and 8 show the PSLL values at different frequencies in Fig. 10 (a) and (b). From the perspective of the PSLR effect, the PSLL of the three arrays in Fig. 10 (a) are stable at low frequency band, and mutate at high frequency band. The lobe levels of DSUE array and DSRE array tend to be stable at high frequency, while the lobe levels of USRE array rise sharply with increasing frequency. This is because the adaptive frequency width of the arrays is related to their non-uniformity. The higher the non-uniformity of aperiodic array is, the more stable its radiation pattern is over a wider frequency range, and the PSLL mutation is reduced. The PSLL of the three array structures in Fig. 10 (b) show similar frequency characteristics as those in Fig. 10 (a). The PSLLs of DSUE array with 4×4 subarrays and USRE array with 7×7 subarrays show a similar tendency as the frequency increases, while the curves of USRE array with 4×4 subarrays and DSUE array with 7×7 subarrays look similar. It is worth mentioning that the PSLR ability of DSRE array with 4×4 subarrays remain stable with the increase of frequency. The PSLL begins to increase at the frequency of 27 GHz which is beyond the frequency range shown in the figure. So, it can be seen that the adaptive frequency width of DSRE array with 4×4 subarrays is very wide. It also has been shown in Fig. 6 that this array has the best PSLR effect. For the arrays proposed in this paper, the higher the non-uniformity of the array, the wider the adaptive frequency width.

## V. CONCLUSION

In this paper, three aperiodic arrangement methods of subarray-level rectangular grid array antennas are proposed. The IBA is used to optimize array structures through the dislocation position of subarrays and random position of elements in the subarray. The PSLR effect of three kinds of arrays with different number of elements and subarrays are analyzed, and the non-uniformity of arrays are characterized by their redundancy. The results show that the PSLR effect of DSRE arrays is better than those of DSUE arrays and USRE arrays. The array with 28×28 elements is taken as an example to study the PSLL in terms of dislocation distance of subarray, scanning angle and applicable frequency. The dislocation distance of subarrays slightly less than one wavelength is the best choice. The DSRE arrays can maintain good PSLR ability at the large scanning angle of up to 60°, which also have wider adaptive frequency width. The proposed design method of DSRE array is a universal and scalable method. For example, the array with 28×28 elements can be treated as a new subarray to further expand the scale of aperiodic array antenna. By inducing the rotation degree of freedom, such as subarray rotation or Fermat spiral, we can combine rotation, dislocation and random elements approaches to study the RSLR effect.

This is of great value to the design and application of large-scale aperiodic array antennas.

## REFERENCES

[1] Caorsi, Salvatore, Andrea Lommi, Andrea Massa and Matteo Pastorino, "Peak sidelobe level reduction with a hybrid approach based on GAs and difference sets," IEEE Transactions on Antennas and Propagation, vol. 52, no. 4, pp. 1116-1121, 2004.

[2] A. B. Smolders and H. J. Visser, "Low Side-Lobe Circularly-Polarized Phased Arrays Using a Random Sequential Rotation Technique," IEEE Transactions on Antennas and Propagation, vol. 62, no. 12, pp. 6476- 6481, Dec. 2014.

[3] Lucas H. Gabrielli and Hugo E. Hernandez-Figueroa, "Aperiodic Antenna Array for Secondary Lobe Suppression," IEEE Photonics Technology Letters, vol. 28, no. 2, pp. 209-212, Jan. 2016.

[4] Thomas G. Spence, Douglas H. Werner, "Design of Broadband Planar Arrays Based on the Optimization of Aperiodic Tilings," IEEE Transactions on Antennas and Propagation, vol. 56, no. 1, pp. 76-86, Jan. 2008.

[5] Randy L. Haupt, "Thinned Arrays Using Genetic Algorithms," IEEE Transactions on Antennas and Propagation, vol. 42, no. 7, pp. 993-999, 1994.

[6] Malik, Hamza, Jehanzeb Burki and Muhammad Zeeshan Mumtaz, "Adaptive Pulse Compression for Sidelobes Reduction in Stretch Processing Based MIMO Radars," IEEE Access, vol. 10, pp. 93231-93244, 2022.

[7] Ghayoula, Elies, Ridha Ghayoula, Jaouhar Fattahi, Emil Pricop, Jean-Yves Chouinard and Ammar Bouallègue, "Radiation pattern synthesis using Hybrid Fourier-woodward-lawson-neural networks for reliable MIMO antenna systems," IEEE International Conference on Systems, Man, and Cybernetics (SMC), pp. 3290-3295, 2017.

[8] Jian Dong, Qingxia Li, Fangmin He, Wei Ni and Yaoting Zhu, "Co-array properties of minimum redundancy linear arrays with minimum sidelobe level," International Symposium on Antennas IEEE, 2008.

[9] Marco Moebus and Abdelhak M. Zoubir, "2D nonuniform array design for imaging applications," International ITG Workshop on Smart Antennas IEEE, pp. 180-183, 2008.

[10] Mustafa Karaman, Ira O. Wygant, Ömer Oralkan and Butrus T. Khuri Yakub, "Minimally Redundant 2-D Array Designs for 3-D Medical Ultrasound Imaging," IEEE Transactions on Medical Imaging, vol. 28, no.7, pp. 1051-1061, 2009.

[11] Yury V. Krivosheev, Alexander V. Shishlov and Vladimir V. Denisenko, "Grating Lobe Suppression in Aperiodic Phased Array Antennas Composed of Periodic Subarrays with Large Element Spacing," IEEE Antennas and Propagation Magazine, vol.57, no. 1, pp. 76-85, Feb.2015.

[12] Hao Wang, Da-Gang Fang and Y. Leonard Chow, "Grating Lobe Reduction in a Phased Array of Limited Scanning," IEEE Transactions on Antennas and Propagation, vol. 56, no.6, pp.1581-1586, Jun. 2008.

[13] Kiersten C. Kerby and Jennifer T. Bernhard, "Sidelobe level and wideband behavior of arrays of random subarrays," IEEE Transactions on Antennas and Propagation, vol.54, no. 8, pp. 2253-2262, Aug. 2006.

[14] Junming Diao, Jakob W. Kunzler and Karl F. Warnick, "Sidelobe Level and Aperture Effifiiciency Optimization for Tiled Aperiodic Array Antennas," IEEE Transactions on Antennas and Propagation, vol. 65, no. 12, pp. 7083-7090, Dec. 2017.

[15] Timothy J. Brockett and Yahya Rahmat-Samii, "Subarray Design Diagnostics for the Suppression of Undesirable Grating Lobes," IEEE Transactions on Antennas and Propagation, vol. 60, no. 3, pp. 1373-1380, Mar. 2012.

[16] Kesong Chen, Xiaohua Yun, Zishu He and Chunlin Han, "Synthesis of Sparse Planar Arrays Using Modified Real Genetic Algorithm," IEEE Transactions on Antennas and Propagation, vol. 55, no. 4, pp. 1067-1073, Apr. 2007.

[17] Robert Macharia Maina, Philip Kibet Lang'at, Peter Kamita Kihato, "Collaborative beamforming in wireless sensor networks using a novel particle swarm optimization algorithm variant," Heliyon, Volume 7, Issue 10, e08247, 2021.

[18] P. Baumgartner, T. Bauernfeind, O. Biro, A. Hackl, C. Magele, W. Renhart and R. Torchio, "Multi-Objective Optimization of Yagi–Uda Antenna Applying Enhanced Firefly Algorithm with Adaptive Cost Function," IEEE Transactions on Magnetics, vol. 54, no. 3, pp. 1-4, March 2018

[19] Heng Liu, Hongwei Zhao, Weimei Li and Bo Liu, "Synthesis of Sparse Planar Arrays Using Matrix Mapping and Differential Evolution," IEEE Antennas and Wireless Propagation Letters, vol. 15, pp. 1905-1908, 2016.

[20] Xiaomin Xu, Cheng Liao, Liang Zhou and Fan Peng, "Grating Lobe Suppression of Non-Uniform Arrays Based on Position Gradient and Sigmoid Function," IEEE Access, vol. 7, pp. 106407-106416, Aug. 2019.

[21] Yusheng Pan and Jiao Zhang, "Synthesis of linear symmetric antenna arrays using improved bat algorithm," Microwave and Optical Technology Letters,vol.62, no. 6, pp. 2383-2389, Dec.2020.

[22] B. Chen, J. Yang, H. Zhang and M. Yang, "An Improved Spherical Vector and Truncated Mean Stabilization Based Bat Algorithm for UAV Path Planning," IEEE Access, vol. 11, pp. 2396-2409, 2023.

[23] Avishek Das, D.Mandal, S.P.Ghoshal and R.Kar, "An efficient side lobe reduction technique considering mutual coupling effect in linear array antenna using BAT algorithm," Swarm and Evolutionary Computation, vol. 35, pp. 26-40, Feb. 2017.

[24] Tao Zheng, Fei Hu, Liang Wu, Jinlong Su and Hao Hu, "Synthesis of Large Alias-Free Field of View Linear Arrays for Synthetic Aperture Interferometric Radiometers," IEEE Transactions on Antennas and Propagation, vol.68, no.12, pp. 7916-7926 , 2020.

[25] Jian Dong, Qingxia Li, Rong Jin, Yaoting Zhu, Quanliang Huangand Liangqi Gui, "A Method for Seeking Low-Redundancy Large Linear Arrays in Aperture Synthesis Microwave Radiometers," IEEE Transactions on Antennas & Propagation, vol. 58, no. 6, pp. 1913-1921, 2010.

[26] Jiao Zhang, Reza Ansari, Xuelei Chen, Jean-Eric Campagne, Christophe Magneville, and Fengquan Wu, "Sky reconstruction from transit visibilities: PAON-4 and Tianlai Dish Array," Monthly Notice of The Royal Astronomical Society, 461, 1950-1966, 2016.

[27] Ching. -Y. Tseng and Lloyd. J. Griffiths, "Sidelobe suppression in minimum redundancy linear arrays," IEEE Sixth SP Workshop on Statistical Signal and Array Processing, 1992.

[28] Jian Dong, Qingxia Li and Wei Guo, "A Combinational Method for Antenna Array Design in Minimum Redundancy MIMO Radars," IEEE Antennas & Wireless Propagation Letters, vol. 8, pp. 1150-1153, 2009.

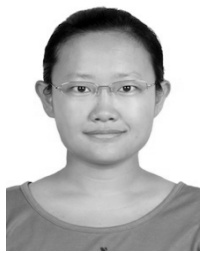

**Jiao Zhang** received the Ph.D. degrees from University of Chinese Academy of Sciences and Universite Paris-Sud, in 2017. Currently she is a lecturer in the School of Physics and Electronic Engineering, Shanxi University, Taiyuan, China. Her current research interests include array signal processing and communication signal processing.

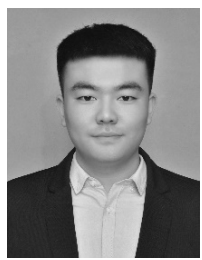

**Hongtao Zhang** was born in Shandong Linyi，China, in 1998. He is now studying for a master's degree in Shanxi University. His research is mainly focused on the synthesis of planar arrays.

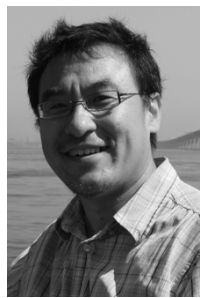

**Xuelei Chen** received the Ph.D. degrees in Physics Department of Columbia University in 1999 and worked as postdoctoral fellow in the Physics Department of Ohio State University and the Institute of theoretical physics of the University of California, Santa Barbara. He joined National Astronomical Observatory at the end of 2004. He mainly engaged in theoretical research and data analysis in cosmology and particle astrophysics. Research topics also include galaxy formation, especially the formation of the first generation of luminous objects in the universe, reionization and their 21 cm observations.

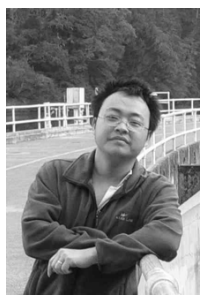

**Fengquan Wu** received the Ph.D. degrees in theoretical physics from Institute of High Energy Physics, Beijing, China, in 2007. Then he joined National Astronomical Observatory, where he worked on the CMB perturbation theory, large scale structure survey and radio astronomy. His current research interests include interferometer array, 21cm radio detection, FRB and transient radio source detection, 21cm global spectrum detection and so on.

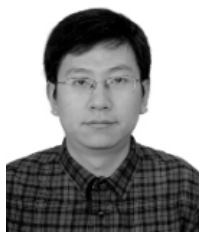

**Yufeng Liu** was born in Shanxi Lvliang, China, in 1986. He received the Ph.D. degree in radio physics from Sichuan University, Sichuan, China, in 2014. In 2015, he joined the School of Physics and Electronic Engineering, Shanxi University. His research is mainly focused on computational electromagnetics and antenna design.

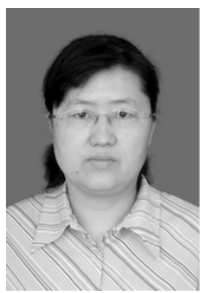

**Wenmei Zhang** received the B.S. and M.S. degrees in electronic engineering from Nanjing University of Science and Technology, Nanjing, China, in 1992 and 1995, respectively, and the Ph.D. degree in electronic engineering from Shanghai Jiao Tong University, Shanghai, China, in 2004. Currently, she is a professor with the School of Physics and Electronic Engineering, Shanxi University, Taiyuan, China. Her research interests include microwave and millimeter-wave integrated circuits, EMC, and microstrip antenna.